\documentclass[prd,twocolumn,showpacs,preprintnumbers,amsmath,amssymb]{revtex4}

\usepackage[utf8]{inputenc}
\usepackage{graphicx}
\usepackage{dcolumn}
\usepackage{bm}
\usepackage{pstricks,pstricks-add}
\usepackage{pst-plot}
\usepackage[all]{xy}
\usepackage[abs]{overpic}
\usepackage{longtable}
\usepackage{subfigure}

\def\be{\begin{equation}}
\def\ee{\end{equation}}

\begin{document}

\title{Test Particle Motion in the Born-Infeld Black Hole}

\author{Rom\'an Linares}
\email{lirr@xanum.uam.mx}
\author{Marco Maceda}
\email{mmac@xanum.uam.mx}
\author{Daniel Mart\'{i}nez-Carbajal}
\email{danielmc@xanum.uam.mx}

\affiliation{Departamento de F\'{\i}sica, Universidad Aut\'onoma Metropolitana Iztapalapa,\\
San Rafael Atlixco 186, C.P. 09340, M\'exico D.F., M\'exico,}

\date{\today}

\begin{abstract}
In this work we review the classification of geodesics of massive test particles in the gravitational background of the black hole solutions of Einstein-Born-Infeld spacetime. Even though some features are quite similar to those of Reissner-Nordström spacetime there are also important differences, particularly those related to the effective potential governing the geodesic motion. Explicit solutions involving Weierstrass functions are given for a pair of specific scenarios. 
\end{abstract}

\maketitle
\section{Introduction}

In 1872 James Clerk Maxwell unified the electricity and the magnetism in a single theory. In Maxwell's theory of electromagnetism, the field of a point-like charge is singular at the position of the charge. As a consequence, it has infinite self-energy. To avoid this unattractive feature, in 1934 Born and Infeld~\cite{Born:1934ji,Born:1934gh} proposed a nonlinear electrodynamics with the goal of obtaining a finite value for the selfenergy of a point-like charge~\cite{Born:1934gh}. In this theory, the electric field of a point charge is regular at the origin. Also, its total energy is finite.

In recent years the Born-Infeld action has received considerable attention due to several reasons. In the context of superstring theory for example, the low energy dynamics of D-branes is governed by the Born-Infeld (BI) action~\cite{Leigh:1989jq}. Also, when analysing the low energy effective action for an open superstring, loop calculations lead to BI type actions~\cite{Fradkin:1985qd}. For detailed discussions on several aspects of the BI theory in string theory see~\cite{Gibbons:2001gy} and~\cite{Tseytlin:1999dj}.

Other motivations arise from a purely constructive generalisation of Einstein-Maxwell systems. In particular the extension of the Reissner-Nordström (RN) black hole solutions in Einstein-Maxwell theory to the charged black hole solutions in EBI theory, with or without a cosmological constant, has attracted some attention in recent years. Different aspects of these black holes have been studied including their thermodynamical properties, phase transitions, geodetical motion and higher-dimensional generalisations~\cite{Wiltshire:1988uq,Rasheed:1997ns,Tamaki:2000ec,Breton:2003tk,Aiello:2004rz,Cataldo:1999wr,Fernando:2003tz,Cai:2004eh,Dey:2004yt}.

In this paper, we will focus on the study of geodesics of electrically and magnetically charged test particles in BI electrodynamics. The black hole solution for EBI gravity was obtained by García et. al.~\cite{Garcia1} in 1984 and Demianski~\cite{Demianski:1986wx} found two years later a static spherically symmetric solution of the EBI equations that is regular at the origin, the so called EBIon. 

The black hole solution we will consider is well known as the nonlinear generalisation of the RN black hole solutions characterised by the mass $M$ and the charge $Q$ of the black hole and the Born-Infeld parameter $b$, that is related to the strength of the electromagnetic field at the position of the charge, usually to be located at the origin. 

In recent years there has been a growing interest in the study of geodesics of certain black holes \cite{Grunau:2010gd,Gibbons:2011rh} and in particular, of the RN solution, which turns out be the ultimate fate of the gravitational collapse of a very massive star charged. In this context, the properties of black hole including its geodesics and its generalisation to non linear electrodynamics is of fundamental interest. Since we already know the black hole solution that generalises the RN solution, it is important to study the complete classification of geodesics for this solution.

There are already some papers written in the literature in this direction; Bretón discussed in a series of papers the test particle trajectories for the static-charged EBI black hole~\cite{Breton:2002td, Breton:2001yk}. Properties of null geodesics of static charged black holes in EBI gravity were presented by Sharmanthie~\cite{Fernando:2012cg} very recently. The aim of our paper is to complete the discussion about the classification of geodesics by analysing the problem in a more systematic way and by addressing some issues that had not been discussed before.

This paper is organised as follows: In Sec.~\ref{secc:2}, we review the EBI solution and discuss the conditions for the existence of an EBI extreme black hole. The geodesic equation for a test particle moving in a EBI spacetime is derived in Sec.~\ref{secc:3} using the Hamilton-Jacobi formalism and the complete classification of the trajectories is presented in Sec.~\ref{secc:4}. Analytic explicit solutions are given in Sec.~\ref{secc:5}, for both the radial and angular differential equations of the geodesic equation. We end up with some remarks in the Conclusions.

Throughout this paper we will use geometrical units $G = c = 1$.

\section{EBI Spacetime }
\label{secc:2}

The story of finding solutions to the Einstein equations of motion coupled to the energy momentum tensor of the  nonlinear electrodynamics of BI~\cite{Born:1934gh}, goes back to the first attempt made by Pellicer and Torrence~\cite{Pellicer:1969cf}. They found a static spherical symmetric solution for a point charge source, which  approaches the RN solution at large distances from the source. 

Some years later Morales~\cite{Morales:1882} found that the Bertotti-Robinson solution admitted an interpretation in terms of nonlinear electrodynamics. Soon after, García et. al.~\cite{Garcia1} found all type-D solutions in the Petrov classification of the EBI system of equations \cite{Garcia1}. Among the solutions they obtained was the generalized RN black hole metric again, usually called EBI black hole. In this section we give a short summary of the way in which the solution is obtained (for a detailed derivation see~\cite{Garcia1}). 

\subsection{EBI black hole}

The action for the gravitational field coupled to a generic nonlinear electrodynamics is 
\begin{equation}
S =\int d^{4}x\sqrt{-g}\left( \frac{R}{16\pi}-{\cal L}(F) \right).
\label{nlaction}
\end{equation}
Here $R$ denotes the curvature scalar obtained from the metric coefficients $g_{\mu\nu}$, $g\equiv\det|g_{\mu\nu}|$
and ${\cal L}(F)$ is the nonlinear electrodynamics Lagrangian density, which depends in a nonlinear way of 
the two invariants of the electromagnetic tensor $F$. For the BI nonlinear electrodynamics we have explicitly
\begin{equation}
\mathcal{L}_{BI}= b^{2}\left(1-\sqrt{1+\frac{F_{\mu\nu}F^{\mu\nu}}{2b^{2}}-
\frac{\left(F_{\mu\nu}\tilde{F}^{\mu\nu}\right)^{2}}{4b^{4}}}\right),
\label{eq:lagrangiana-N-L}
\end{equation}
where 
\be
\tilde{F}_{\mu\nu} = -\frac{\epsilon_{\mu\nu\rho\sigma}}{2\sqrt{-g}}F^{\rho\sigma}, 
\ee
denotes the dual tensor of the electromagnetic tensor and $\epsilon_{\mu\nu\rho\sigma}$ is the totally antisymmetric Levi-Civita 
tensor. The parameter $b$ is the maximum electromagnetic field intensity and has dimensions of [length]$^{-2}$. Notice that this 
Lagrangian reduces to the Maxwell one in the strong field limit ($b\rightarrow\infty$)
\begin{equation}\label{MaxLimit}
{\cal L}_{BI}(F)=-\frac14 F_{\mu\nu}F^{\mu\nu}+\mathcal{O}(F^{4}).
\end{equation}

The full system of equations of motion derived from the action Eq.~(\ref{nlaction}) is given by the Einstein field equations 
\begin{equation}
R_{\mu\nu}-\dfrac{1}{2}g_{\mu\nu}R=8\pi T_{\mu\nu}, 
\label{eq:tensor de Einstein}
\end{equation}
and the electromagnetic field equations
\begin{equation}
\nabla_{\mu}\left(F^{\mu\nu}{\cal L},_{F}\right)=0. \label{eq:ecuaciones de campo}
\end{equation}
In the field equations, Eq.~(\ref{eq:tensor de Einstein}), the energy momentum tensor is given by
\begin{equation}
T_{\mu\nu}={\cal L}_{BI} \, g_{\mu\nu}-F_{\mu\sigma}F_{\;\nu}^{\sigma}, 
\label{eq:T energ-momento}
\end{equation}
and in the conservation laws, Eq.~(\ref{eq:ecuaciones de campo}), ${\cal L},_{F}$ represents the partial derivative of ${\cal L}_{BI}(F)$ with respect to $F$.

The static electrically charged black hole solution with spherical symmetry for the EBI system of equations is well-known, it is given by the metric
\begin{equation}
ds^{2}=- \Delta dt^{2}+\dfrac{dr^{2}}{\Delta} + r^{2}(d\theta^{2}+\sin^{2}\theta d\varphi^{2}),
\label{eq:B-I}
\end{equation}
and  the radial electric field
\begin{equation}
F_{\mu\nu}= \frac{Q}{\sqrt{r^{4}+Q^{2}/b^{2}}} \,(\delta_{\mu}^{r}\delta_{\nu}^{t}-\delta_{\mu}^{t}\delta_{\nu}^{r}).
\label{Efield}
\end{equation}
The function $\Delta = \Delta(r)$ in the metric Eq.~(\ref{eq:B-I}) is given by
\begin{eqnarray}
\Delta &=& 1-\frac{2M}{r}+\dfrac{2}{3}b^{2}r^{2}\left(1-\sqrt{1+Q^{2}/b^{2}r^{4}}\right) 
\nonumber \\[4pt]
&&+ \dfrac{4Q^{2}}{3r} \intop_{r}^{\infty}\dfrac{ds}{\sqrt{s^{4}+Q^{2}/b^{2}}}.
\label{eq:gtt B-I}
\end{eqnarray}
The last term is an elliptic integral of the first kind, which in the literature can be found written either in terms of the Legendre's elliptic integral: $F(\beta,\,\kappa)\equiv\int_{\beta}^{\infty}(1-k^{2}\sin^{2}s)^{-1/2} \, ds$, or in terms of the hypergeometric function ${}_2F_{1} (a, b; c; x)$ as follows
\begin{eqnarray}
\intop_{r}^{\infty}\dfrac{ds}{\sqrt{s^{4}+Q^{2}/b^{2}}} &=& 
\dfrac{1}{2}\sqrt{\dfrac{b}{Q}}F\left[ \arccos\left( \tfrac {br^{2}/Q - 1}{br^{2}/Q + 1} \right) ,\,\tfrac{1}{\sqrt{2}} \right]
\label{eq:INT-5-1}
\nonumber \\[4pt]
& = & \frac{1}{r}\,_{2}F_{1}\left(\frac{1}{4},\,\frac{1}{2};\,\frac{5}{4};-\frac{Q^{2}}{b^{2}r^{4}}\right).
\label{eq:hypergeometric}
\end{eqnarray}
For a detailed deduction of Eq.~(\ref{eq:hypergeometric}) see~\cite{Gunasekaran:2012dq}. 

In this paper we have chosen to follow~\cite{Breton:2002td} and write down the function $\Delta$ in terms of the Legendre's elliptic function
\begin{eqnarray}
\Delta &=& 1-\frac{2M}{r}+\dfrac{2}{3}b^{2}r^{2}\left(1-\sqrt{1+Q^{2}/b^{2}r^{4}}\right)
\nonumber \\[4pt]
&&+ \dfrac{2 Q^{2}}{3r}\,\sqrt{\dfrac{b}{Q}} F\left[ \arccos\left( \tfrac{br^{2}/Q-1}{br^{2}/Q+1} \right) ,\,\tfrac{1}{\sqrt{2}} \right].   \label{eq:gtt B-I-2}
\end{eqnarray}
The physical interpretation of the parameters in the function $\Delta$ is the following: $M$ is the mass and $Q$ is the electric charge of the black hole. $b$ is the BI parameter which corresponds to the magnitude of the electric field at $r = 0$. The solution can have either zero (naked singularity), one or two horizons depending on the values of these parameters. This conclusion is obtained by simple inspection of the condition $\Delta=0$.  

\begin{center}
\begin{figure}[pbth]
\includegraphics[width=8cm]{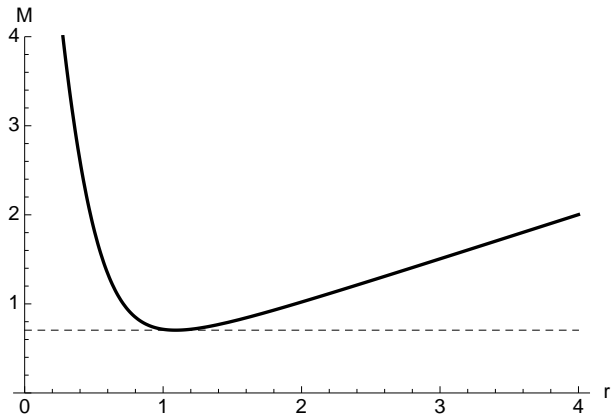}
\caption{The mass $M$ of the EBI black hole as a function of $r$ ($Q = 2, b = 5$); at $M = 0.7$ there is only one horizon.}  
\label{fig1}
\end{figure}
\end{center}

To have a better understanding on the nature of the horizons, we have plotted in Fig.~\ref{fig1} the mass $M$ as a function of the horizon radius. For the sake of clarity we have fixed the value of the BI parameter $b$ and the value of the electric charge $Q$ as well.  

As can be seen from this plot, there is a critical value $M^\star$ for the mass of the black hole that leads to different physical scenarios: first, for values of $M < M^\star$ we have a naked singularity; we will not discuss this case any further in this paper. For $M = M^\star$ we have a black hole solution with one horizon (dashed line) and for values of $M > M^\star$ we have a black hole with two horizons; we will denote by $r_{h_\pm}$ the inner (outer) radii respectively in this case.

The metric in Eq.~(\ref{eq:B-I}) has the expected limits. In the  strong field limit, $b\rightarrow\infty$, we recover the RN black hole solution in agreement with the Maxwell limit Eq.~(\ref{MaxLimit}). As expected in this limit, the radial electric field (\ref{Efield}) approaches the Maxwellian expression of the electric field $E = Q/r$, which diverges at the origin. As for the function $\Delta$ we have 
\begin{equation}
\lim_{b \rightarrow \infty} \Delta = 1 - \frac{2M}{r} + \frac{Q^2}{r^2}.
\end{equation}
As usual, by setting further $Q = 0$, we obtain also the Schwarzschild black hole solution. Additionally, we can obtain the Schwarzschild black hole solution by taking the weak field limit $b \rightarrow 0$ and then $Q = 0$. 

For large values of $r$, $r \rightarrow \infty$ with $b\neq 0$ and finite, the function $\Delta$ becomes the unity and we obtain a flat metric, meaning that the EBI black hole is an asymptotically flat solution. It is clear that for small values of the $b$ parameter ($b \ll 1$) we have a black hole solution that looks like very similar to the Schwarzschild black hole and for large values of $b$ ($b \gg 1$) we have a solution that is very similar to the RN black hole.

As has been showed in \cite{Garcia1}, because the BI theory has the freedom of electromagnetic duality rotations, the EBI black hole solution can include also a magnetic charge $G$; the corresponding solution is obtained simply from the electric charged case by the substitution $Q\rightarrow\sqrt{Q^{2}+G^{2}}$.

\subsection{Extreme Black Holes}

The necessary and sufficient conditions to have an extreme EBI black hole solution ($r_{h_+} = r_{h_-} \equiv r_{ex}$) are $\Delta =0$ and $d\Delta/dr=0$. Combining both conditions, we obtain from Eq. (\ref{eq:gtt B-I-2}) a constraint that determines the horizon radii $r_{ex}$ for the extreme black hole in terms of the electric charge
\begin{equation}
1+2\left(b^2\, r_{ex}^{2}-\sqrt{b^4\, r_{ex}^{4}-Q^{2}b^{2}}\right)=0, 
\end{equation}
its solution being given by
\begin{equation}
r_{ex}^{2}=Q^{2}-\frac{1}{4b^{2}}.
\label{eq:Horizonte-extremo}
\end{equation}
Hence the horizon belonging to the extreme EBI black hole is determined by the positive root of Eq.~(\ref{eq:Horizonte-extremo}), i.e., 
\be
r_{ex}=\sqrt{Q^{2}-\frac{1}{4b^{2}}}.
\ee 
This solution is meaningful only if the radicand is positive, i.e. if $Q>1/2b$. When the radicand is zero, we have the case of a spacetime singularity. In the case $Q<1/2b$ we obtain a naked singularity.

\begin{center}
\begin{figure}[pbth]
\includegraphics[width=8cm]{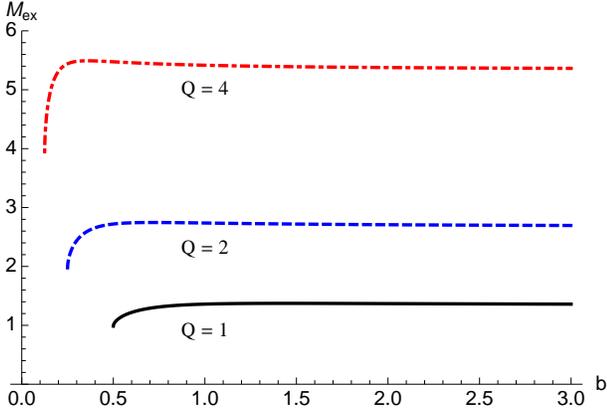}
\caption{The mass $M_{ex}$ of the extreme EBI black hole as a function of $b$.}  
\label{fig2}
\end{figure}
\end{center}

It is possible to express the extremality condition as a function of the form $M=M(b\,,Q\,)$ by substituting back the expression of $r_{ex}$ in the condition $\Delta=0$. This gives
\begin{eqnarray}
M(r_{ex}) &=& \frac{r_{ex}}{2}-\frac{b^{2}r_{ex}^{3}}{3}\left(1-\sqrt{1+Q^{2}/b^{2}r_{ex}^{4}}\right) 
\nonumber \\[4pt]
&&+ \dfrac{Q^{2}}{3}\sqrt{\dfrac{b}{Q}}F\left(\arccos\left\{ \tfrac{br_{ex}^{2}/Q-1}{br_{ex}^{2}/Q+1}\right\} ,\,\tfrac{1}{\sqrt{2}}\right). 
\label{eq:Masa-extremo}
\end{eqnarray}
In the strong field limit, $b \to \infty$, this condition reduces to the well-known condition for the extremal RN black hole solution $r_{ex}=M=Q$. Fig.~\ref{fig2} shows $M(r_{ex})$ as a function of $b$ for $Q$ fixed. It is clear that for a given value of $Q$, the horizon size depends clearly upon the choice of $b$.

\section{The geodesic equation}
\label{secc:3}

The EBI solution is described by the metric in Eq.~(\ref{eq:B-I}). In the following we will be concerned with the dyon case where both the electric charge $Q$ and the magnetic charge $G$ are nonvanishing. In this scenario, the field strength $F_{\mu\nu}=A_{\nu,\mu}-A_{\mu,\nu}$, and its dual $\tilde{F}_{\mu\nu}=\tilde{A}_{\nu,\mu}-\tilde{A}_{\mu,\nu}$ are derived respectively from the vector potentials $A_{\mu}$ and $\tilde {A}_{\mu}$; their explicit forms are known from the analysis in~\cite{Garcia1} of type-D solutions in EBI spacetime 
\begin{eqnarray}
A_{t}=Q\intop_{r}^{\infty}\dfrac{ds}{\sqrt{s^{4}+Q^{2}/b^{2}}}, & \quad & A_{\phi}=-G\cos\theta,
\nonumber \\[4pt]
\tilde{A}_{t}=iG\intop_{r}^{\infty}\dfrac{ds}{\sqrt{s^{4}+Q^{2}/b^{2}}}, &  & \tilde{A}_{\phi}=iQ\cos\theta.
\label{vectpotbi}
\end{eqnarray}

The geodesic motion of test particles in the EBI spacetime can be analysed using the Hamilton-Jacobi (HJ) equation, which can be constructed from a constant of motion that we always have at our disposal for geodesics: metric compatibility implies that along the path the quantity 
\be
\delta=-g_{\mu\nu}\frac{dx^{\mu}}{d\lambda}\frac{dx^{\nu}}{d\lambda}
\ee
is constant. Of course, for a massive particle we typically choose $\lambda=\tau$ (proper time), $\delta=1$, and the above relation simply becomes $m^{2}\delta=-g^{\mu\nu}p_{\mu}p_{\nu}$. For a massless particle we always have $\delta=0$. We will also be concerned with spacelike geodesics (even though they do not correspond to paths of particles), for which we will choose $\delta=-1$. 

The Hamilton-Jacobi equation is given by~\cite{Landau:1979tf} 
\begin{equation}
m^{2}\delta=-g^{\mu\nu}\left(\dfrac{\partial S}{\partial x^{\mu}}\right)\left(\dfrac{\partial S}{\partial x^{\nu}}\right).
\label{eq:H-J final}
\end{equation}
Using the minimal coupling defined by $p_{\mu}\rightarrow p_{\mu}-qA_{\mu}+ig\check{A}_{\mu}$ to account for all 
electromagnetic interactions, the HJ equation for a particle with electric charge $q$ and magnetic 
charge $g$ is
\begin{equation}
m^{2}\delta=-g^{\mu\nu}\left(\dfrac{\partial S}{\partial x^{\mu}}-qA_{\mu} + 
ig\check{A}_{\mu}\right)\left(\dfrac{\partial S}{\partial x^{\nu}}-qA_{\nu}+ig\check{A}_{\nu}\right).
\label{eq:H-J 1}
\end{equation}
In our case, the Hamiltonian does not depend explicitly on the coordinates $\tau$ and $\phi$, i.e., these coordinates are cyclical and thus there are conserved quantities. This allows us to consider the following Ansatz 
\be
S = -E t + L\phi+S_1(r)+S_2(\theta),
\label{ansatzs}
\ee
for the action $S$. $\delta$, as mentioned before, is equal to $0$ for a massless particle and equal to $1$ for a massive particle. On the other hand, the constants $E$ and $L$ are identified respectively with the energy and the angular momentum, along the $z$ direction, of the particle. As noted from Eq.~(\ref{vectpotbi}), the terms $qA_{\phi}$ and $q\check{A}_{\phi}$ in Eq.~(\ref{eq:H-J 1}) have a non-trivial dependence on the angular variable $\theta$; in consequence, even though the EBI metric Eq.~(\ref{eq:B-I}) is spherically symmetric, the motion followed by a massive test particle possesses axial symmetry.

There are two Killing vectors associated with the stationarity and axisymmetry of the EBI spacetime: 
\begin{eqnarray}
&&\xi_{(t)}^{\mu}\equiv(\partial_{t})^{\mu}=(1,\,0,\,0,\,0),
\nonumber \\[4pt]
&&\xi_{(\phi)}^{\mu}\equiv(\partial_{\phi})^{\mu}=(0,\,0,\,0,\,1).
\end{eqnarray}
The EBI spacetime also has an irreducible Killing tensor given by
\begin{equation}
K_{\mu\nu}\equiv2r^{2}l_{(\mu}n_{\nu)}+r^{2}g_{\mu\nu}=2r^{2}m_{(\mu}\bar{m}_{\nu)},
\end{equation}
with the null tetrad defined by 
\begin{eqnarray}
l^{\mu} &\equiv& (r^{2},\,\Delta_r,\,0,\,0)/\Delta_r
\nonumber \\[4pt]
n^{\mu} &\equiv& (r^{2},-\Delta_r,\,0,\,0)/2r^{2}, 
\nonumber \\[4pt]
m^{\mu} &\equiv& (0,\,0,\,1,\, i/\sin\theta)/\sqrt{2}r,
\end{eqnarray}
where $l^{\mu}n_{\mu}=-1$ and $m^{\mu}\bar{m}_{\mu}=1$ while all the other inner products vanish. The metric $g_{\mu\nu}$ can be written
in terms of the null vectors as $g_{\mu\nu}=-2l_{(\mu}n_{\nu)}+2m_{(\mu}\bar{m}_{\nu)}$. Here we have defined $\Delta_r := r^2 \Delta$.

Leaving aside the effect of self-interaction, a particle in EBI spacetime can be regarded as a test particle that moves along a geodesic; its coordinates $t(\tau),\, r(\tau),\,\theta(\tau)$ and $\phi(\tau)$ are parametrised by the proper time $\tau$. Furthermore, there are three integrals of motion from the symmetries of the EBI spacetime: the energy $E$, angular momentum $L$ and Carter constant $K$~\cite{Carter:1968rr}, respectively. These are expressed as
\begin{eqnarray}
E & \equiv & -\xi_{(t)}^{\mu}p_{\mu}=m\dfrac{\Delta_r}{r^{2}}\dfrac{\partial t}{\partial\tau}+\Delta_{q}I(r),
\nonumber \\[4pt]
L & \equiv & \xi_{(\phi)}^{\mu}p_{\mu}=mr^{2}\sin^{2}\theta\dfrac{\partial\phi}{\partial\tau}-\Delta_{g}\cos\theta,
\nonumber \\[4pt]
K & \equiv & K^{\mu\nu}p_{\mu}p_{\mu}=p_{\theta}^{2}+L^{2}\left(\frac{\cos\theta}{\sin\theta}\right)^{2}.
\label{eq:Energ=0000EDa}
\end{eqnarray}
Here $I(r)$ is the integral in Eq.~(\ref{eq:hypergeometric}). Because functions of conserved quantities are also conserved, any function of $K$ and the two other constants of the motion can be used as a third constant in place of $K$. This results in some confusion as to the form of Carter's constant. For example, it is sometimes more convenient to use $k := K+L^{2}$ as the conserved quantity of motion.

For later convenience, we define dimensionless quantities ($r_s :=2M$)
\begin{eqnarray}
&&\tilde{r} :=\dfrac{r}{r_{s}},\quad\tilde{t} :=\dfrac{t}{r_{s}},\quad\tilde{\tau} :=\dfrac{\tau}{r_{s}},
\nonumber \\[4pt]
&&\tilde{Q} :=\dfrac{Q}{r_{s}},\quad\tilde{G} :=\dfrac{G}{r_{s}},\quad\tilde{L} :=\dfrac{L}{r_{s}}.
\end{eqnarray}

The use of the Ansatz Eq.~(\ref{ansatzs}) in the HJ equation leads to a differential equation for each coordinate. At this stage, it is more convenient to parametrize the particle orbit with the so-called \emph{Mino time} $\gamma$, which is related to the parameter $\tilde \tau$ as $d\tilde \tau \equiv \tilde r^{2}\, d\gamma$~\cite{Mino:2005an}. In terms of the Mino time, the first set of geodesic equations of motion are 
\be
\left(\dfrac{d\tilde{r}}{d\gamma}\right)^{2} = R, \quad \left(\dfrac{d\theta}{d\gamma}\right)^{2} = \Theta,
\label{eq:R-Geod=0000E9sica 0}
\ee
where
\begin{eqnarray}
&&R := \dfrac{\tilde{r}^{4}}{m^{2}}\left[ E+\Delta_{q}I(\tilde r) \right]^{2}-\dfrac{\tilde{\Delta}_r}{m^{2}}(m^{2}\delta\tilde{r}^{2}+k),
\label{eq:R-Geod=0000E9sica 1}
\\[4pt]
&&\Theta := \dfrac{k}{m^{2}}-\dfrac{1}{m^{2}\sin^{2}\theta}(\tilde{L}+\Delta_{g}\cos\theta)^{2},
\label{eq:Theta-Geod=0000E9sica 2-2}
\end{eqnarray}
meanwhile the second set is
\begin{eqnarray}
&&\dfrac{d\phi}{d\gamma} = \dfrac{1}{m\,\sin^{2}\theta}(\tilde{L}+\Delta_{g}\cos\theta),
\label{eq:Geod=0000E9sica-L}
\\[4pt]
&&\dfrac{d\tilde{t}}{d\gamma} = \dfrac{\tilde{r}^{4}}{m\,\tilde{\Delta}_{r}}\left[ E-\Delta_{q}I(\tilde{r}) \right].
\label{eq:Geod=0000E9sica-E}
\end{eqnarray}
In the above expressions the following notations were used~\cite{Grunau:2010gd}: $\Delta_{g}\equiv\tilde{G}q - \tilde{Q}g$, $\Delta_{q}\equiv\tilde{Q}q + \tilde{G}g\,$, $\tilde{\Delta}_r \equiv \Delta_r/r_{s}^{2}$ and $I(\tilde r)$ is the integral in Eq.~(\ref{eq:hypergeometric}) after rescaling.

\section{Classification of geodesics }
\label{secc:4}

We now proceed to solve the Hamilton-Jacobi Eqs.~(\ref{eq:R-Geod=0000E9sica 0})-(\ref{eq:Geod=0000E9sica-E}). They are rather complicated due to the polynomial $R$ in Eq.~(\ref{eq:R-Geod=0000E9sica 1}) and the function $\Theta$  in Eq.(\ref{eq:Theta-Geod=0000E9sica 2-2}). These polynomials depend strongly on the constants of motion, the metric coefficients and the charges of the test particle and this in turn will influence the possible types of orbits that a particle may follow. To begin with, in this section we discuss the charged motion in the EBI spacetime.
    
\subsection{The $\theta-$motion}

The polar angle $\theta$ should certainly take only real values. From Eq.~(\ref{eq:Theta-Geod=0000E9sica 2-2}), we see that real solutions are allowed if the condition $\Theta\geq0$ holds. This means that $k \geq 0$. Using now the new variable $\xi := \cos\theta$, Eq.~(\ref{eq:Theta-Geod=0000E9sica 2-2}) becomes 
\begin{eqnarray}
\left(\dfrac{d\xi}{d\gamma}\right)^{2}=\Theta_{\xi} & \textrm{with} & \Theta_{\xi} = a\xi^{2}+b\xi+c,
\label{eq:Geodesic-cos-Theta}
\end{eqnarray}
where
\be
a=-(k+\Delta_{g}^{2}), \qquad b=-2\tilde{L}\Delta_{g},  \qquad c=k-\tilde{L}^{2}. 
\ee
It should be noticed that we obtain a quadratic polynomial on the right hand side of this equation. It follows $a<0$ since $k \geq 0$. The turning points where $\Theta_{\xi}$ vanishes define the angles of two cones and the motion of test particles is confined to this region; it has been pointed out before that a similar feature appears in Taub-NUT and Kerr spacetimes~\cite{Kagramanova:2010bk, Hackmann:2010zz}.  
In the special case when $\Delta_{g}$ vanishes, the motion takes place on a plane, as exemplified by the orbits of only electrically charged or neutral particles in a RN spacetime.

Let us now focus on the requirement $\Theta_\xi \geq0$.  We have first that the zeroes of this polynomial are given by 
\be
\xi_{1,2} = -\frac{\tilde{L}\Delta_{g}\pm\sqrt{k\kappa}}{k+\Delta_{g}^{2}},
\ee
where $\kappa := k - \tilde{L}^{2} + \Delta_{g}^{2}$. Since $k \geq 0$, for these zeroes to be real we must have $\kappa \geq 0$. It can be easily seen that $\xi \in [-1,1]$ and $\Theta_{\xi} \geq 0$ are then guaranteed.

On the other hand, the maximum of $\Theta_\xi$ is at $(-\frac{\tilde{L}\Delta_{g}}{k+\Delta_{g}^{2}},\:\frac{k\kappa}{k+\Delta_{g}^{2}})$. If $\tilde{L}$ and $\Delta_{g}$ were vanishing, then the zeroes would be symmetric with respect to the line $\xi=0$. Physically this means that only for vanishing $\tilde{L}$ or $\Delta_{g}$, the motion has a symmetry with respect to the equatorial plane. The essential features of the orbits of the $\theta-$motion can then be classified in quite a similar way as that given in~\cite{Grunau:2010gd} for the RN spacetime and we shall not dwell on this.

\subsection{The $\tilde r$-motion}

We now explore the dynamics on the $\tilde r$-coordinate for massive particles and shall start as for the $\theta$-motion, namely, we require real values for $\tilde r$.  Clearly this implies $R \ge 0$. Now, the regions where this condition is satisfied are bounded by the zeroes of $R$ and we can further analyse these regions by looking for roots of multiplicity 2 of the function $R$. More specifically we consider the conditions\begin{eqnarray}
&&R =  \dfrac{\tilde{r}^{4}}{m^{2}}\left[ E+\Delta_{q}I(\tilde r) \right]^{2}-\dfrac{\tilde{\Delta}_r}{m^{2}}(m^{2}\delta\tilde{r}^{2}+k) = 0, 
\nonumber \\[4pt]
&&\frac {dR}{d\tilde r} = 0.
\end{eqnarray} 
In a similar way as for the RN case~\cite{Grunau:2010gd}, parametric plots on the $(E, k)$-plane can be done. For comparison purposes, we focus only on the $Q = 0.3, G = 0.1, q = 0.1, g = 0$ situation for different values of the BI parameter $b$. 

\begin{center}
\begin{figure}[pbth]
\includegraphics[width=8cm]{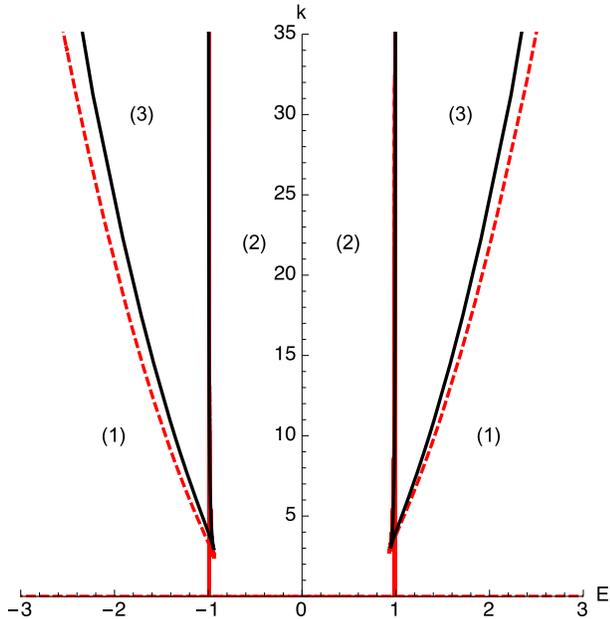}
\caption{Distribution of roots of the function $R$ on the $k-E$ plane showing the transition from Schwarzschild spacetime ($\tilde b=1\times 10^{-6}$, solid line) to RN spacetime ($\tilde b = 1\times 10^6$, dashed line); the region with 4 zeroes is not indicated.}  
\label{fig3}
\end{figure}
\end{center}

As we see in Fig.~\ref{fig3}, for $b$ large we recover the RN curves (dashed line) and for $b$ small the Schwarzschild limit; note for the latter a more symmetric distribution of the common zeroes of $R$ and $dR/d\tilde r$. The general features of these curves are similar to the RN case discussed in~\cite{Grunau:2010gd}.

\begin{center}
\begin{figure}[pbth]
 \includegraphics[width=8cm]{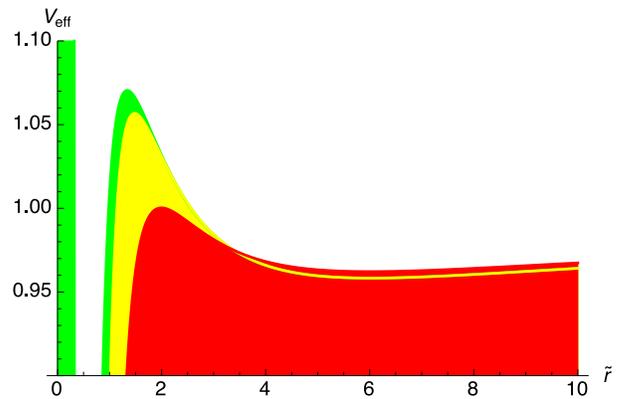}
\caption{$\tilde Q = 0.4, \tilde G = 0.25, \tilde q = 0.05, \tilde g = 0.1, k = 4$}  
\label{fig7}
\end{figure}
\end{center}

\begin{center}
\begin{figure}[pbth]
 \includegraphics[width=8cm]{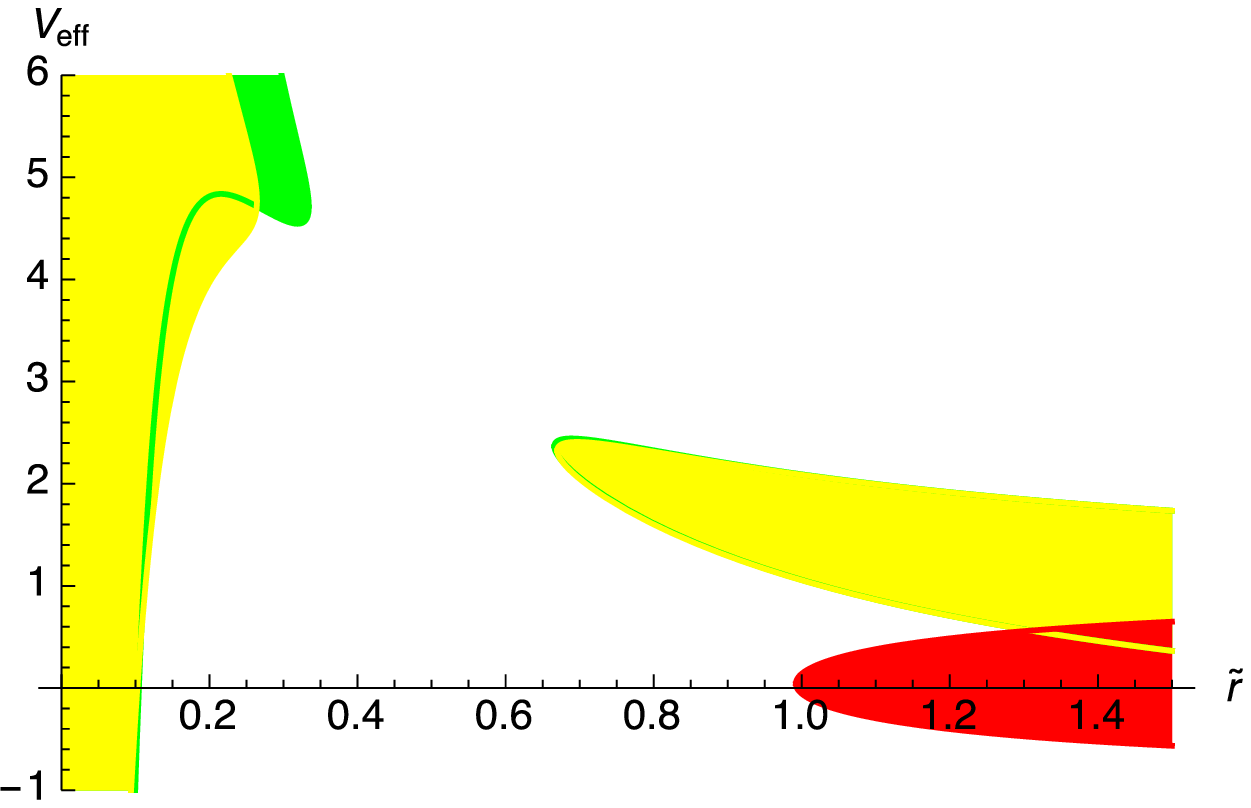}
\caption{$\tilde Q = 0.4, \tilde G = 0.25, \tilde q = -4, \tilde g = 0.1, k =0.2$}  
\label{fig8}
\end{figure}
\end{center}

\begin{center}
\begin{figure}[pbth]
 \includegraphics[width=8cm]{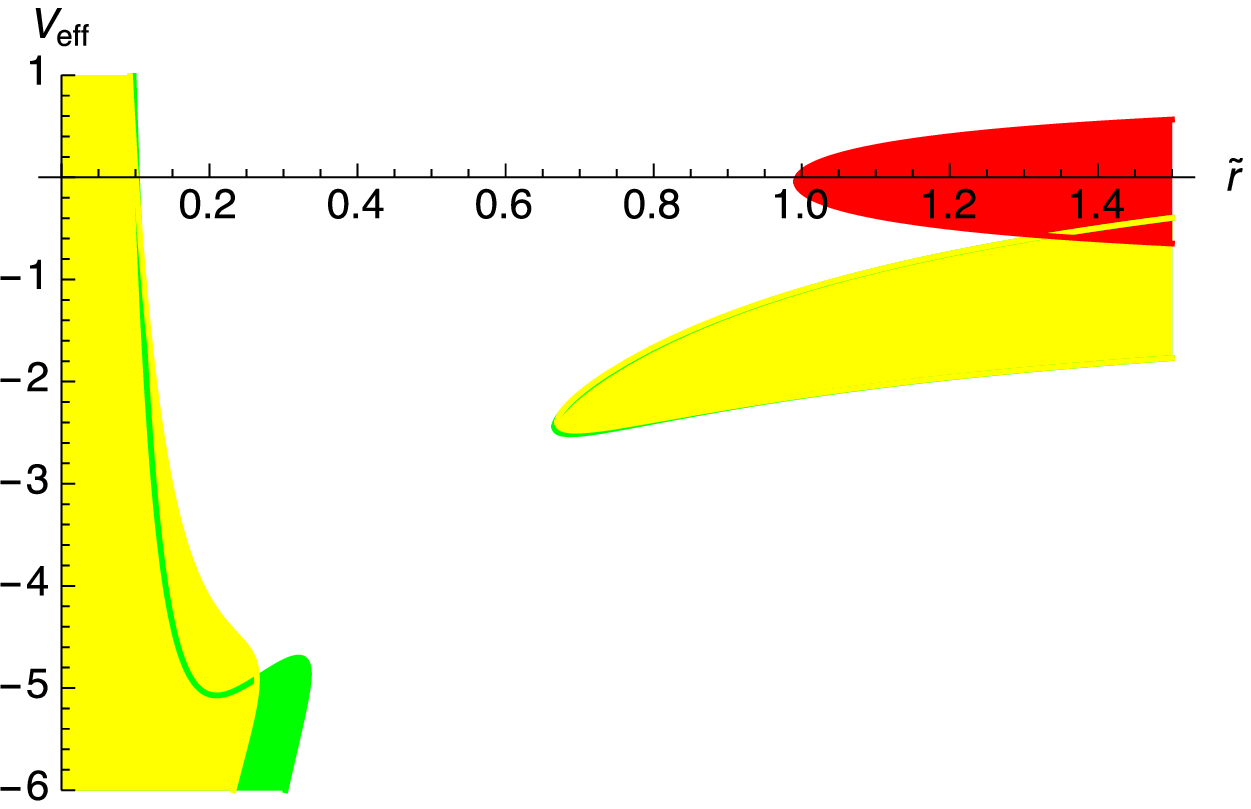}
\caption{$\tilde Q = 0.4, \tilde G = 0.25, \tilde q =4, \tilde g = 0.1, k = 0.2$}  
\label{fig9}
\end{figure}
\end{center}

\begin{center}
\begin{figure}[pbth]
 \includegraphics[width=8cm]{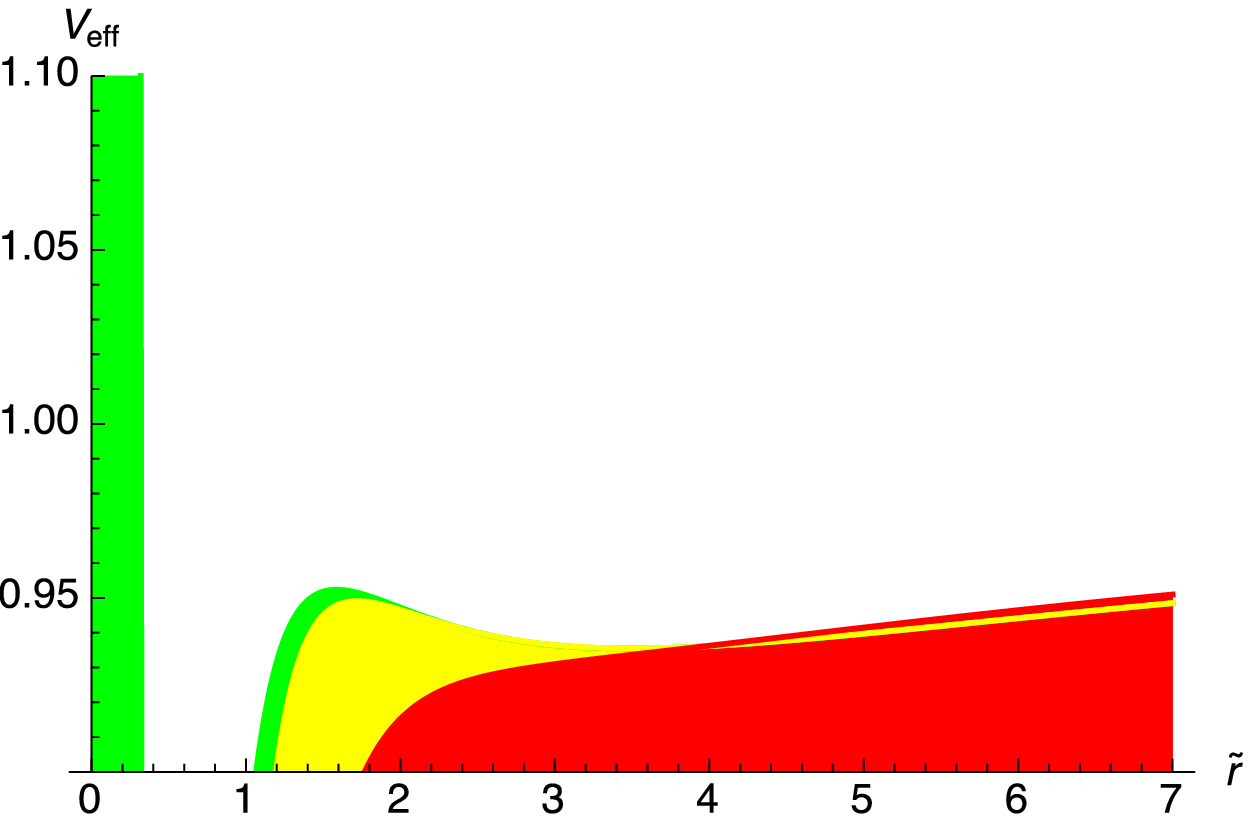}
\caption{$\tilde Q=0.4, \tilde G= 0.25, \tilde q= 0.025, \tilde g= 0.1, k=2.7$}  
\label{fig10}
\end{figure}
\end{center}

Furthermore, along the lines of~\cite{Grunau:2010gd}, we can determine the turning points of the orbits followed by massive particles. From Eq.~(\ref{eq:R-Geod=0000E9sica 0}) the constraint
\begin{equation}
0=\left(\dfrac{d\tilde{r}}{d\gamma}\right)^{2}=\tilde{r}^{4}(E-V_{eff}^{+})(E-V_{eff}^{-}).
\label{eq:condicion de potencialefectivo}
\end{equation}
defines an effective potential of the form
\begin{equation}
V_{eff}^{\pm}=-\Delta_{q}I(\tilde{r})\pm\dfrac{1}{\tilde{r}^{2}}\sqrt{\tilde{\Delta}_{r}(\delta\tilde{r}^{2}+k)}.
\end{equation}
In Figs.~\ref{fig7}-\ref{fig10} we show this potential for some values of the parameters $\tilde Q, \tilde G, \tilde q, \tilde g, k$ and the BI parameter $\tilde b$. The red area shows the Schwarzschild limit ($\tilde b \ll 1$), the green are corresponds to the RN limit ($\tilde b \gg 1$) and the yellow region is associated to a generic EBI case ($\tilde b \sim 0.1 - 3$). It can be remarked the absence of a barrier wall near the origin for EBI in Figs.~\ref{fig7} and~\ref{fig10}.

\vspace{1cm}

\section{Solution of the geodesic equation}
\label{secc:5}

We now proceed to discuss some analytical solutions to the equations of motion Eqs.~(\ref{eq:R-Geod=0000E9sica 0})\textendash{}(\ref{eq:Geod=0000E9sica-E}). 

\subsection{Solution of the $\tilde r$-equation of motion}

Given the complexity of Eq.~(\ref{eq:R-Geod=0000E9sica 1}), analytical solutions in terms of elementary functions are not known, however there are cases for which it is possible to simplify the equations in such a way that explicit analytical solutions can be found, in particular, there are two limit cases where explicit details can be worked out.

\subsubsection{Case $Q / b \gg r^2$}

The function $\Delta_r$ has the expression
\be
\Delta_{r} =r^{2}-2Mr+2b^{2}r\intop_{r}^{\infty}ds\left(\sqrt{s^{4}+Q^{2}/b^{2}}-s^{2}\right).
\ee
Using now the relation~\cite{Gunasekaran:2012dq}
\begin{widetext}
\be
\int_{r}^{\infty}(\sqrt{s^{4}+r_{0}^{4}}-s^{2})ds = \frac{1}{3}r^{3}\left(1-\sqrt{1+r_{0}^{4}/r^{4}}\right) + \frac{2}{3}r_{0}^{4}\int_{r}^{\infty}\frac{ds}{\sqrt{s^{4}+r_{0}^{4}}} \simeq \frac{r_{0}^{3}}{6\sqrt{\pi}}\Gamma\left(\frac{1}{4}\right)^{2}-r_{0}^{2}\, r - \frac{1}{10}\frac{r^{5}}{r_{0}^{2}}+\mathcal{O}(r^{9}),
\ee
\end{widetext}
with $r_0^4 := Q^2/b^2$, we obtain
\be
\Delta_{r}  =  -2(M-M_{m})r+(1-2bQ)r^{2}+\mathcal{O}(r^{6}),
\ee
where we have defined
\be
M_{m} := \frac{1}{6}\sqrt{\frac{b}{\pi}}Q^{3/2}\Gamma\left(\frac{1}{4}\right)^{2}.
\label{eq:Delta-INT-1-1-1-1-1}
\ee
It follows that
\begin{equation}
\tilde{\Delta}_{r}=\tilde{r}(\tilde{M}-1)+(1-2\tilde{b}\tilde{Q})\tilde{r}^{2},
\label{eq:gtt b peque=00003D0000F1o}
\end{equation}
with $\tilde{M}=M_{m}/2M$. Let us now define 
\be
b_{1} := \tilde{M}-1, \qquad b_{2} := 1-2\tilde{b}\tilde{Q},
\ee
then the $\tilde r$-equation of motion becomes
\begin{equation}
\left(\dfrac{\partial\tilde{r}}{\partial\gamma}\right)^{2}=\tilde{r}^{4}\left(E+\frac{\Delta_{q}}{\tilde{r}}\right)^{2}-(b_{1}\tilde{r}+b_{2}\tilde{r}^{2})(\delta\tilde{r}^{2}+k),
\end{equation}
or equivalently
\begin{equation}
\left(\dfrac{\partial\tilde{r}}{\partial\gamma}\right)^{2} = a_{1}\tilde{r}+a_{2}\tilde{r}^{2}+a_{3}\tilde{r}^{3}+a_{4}\tilde{r}^{4},
\label{eq:Ec cuarto orden}
\end{equation}
where
\begin{eqnarray}
&&a_{1}=-b_{1}k,\qquad\qquad a_{2}=\Delta_{q}^{2}-b_{2}k,
\nonumber \\[4pt]
&&a_{3}=2\Delta_{q}E-b_{1}\delta,\qquad a_{4}=E^{2}-b_{2}\delta.
\end{eqnarray}
By making the change of variable $\tilde{r}=1/x$, Eq~.(\ref{eq:Ec cuarto orden}) can be cast as the differential equation
\begin{equation}
\left(\dfrac{\partial x}{\partial\gamma}\right)^{2}=a_{1}x^{3}+a_{2}x^{2}+a_{3}x+a_{4}.
\end{equation}
Notice that the polynomial on the right-hand side of this equation is of third order on $x$. Finally, through the defining relation
\be
x = 4y/a_{1}-a_{2}/3a_{1},
\ee
we obtain the standard form of the Weierstrass differential equation
\begin{equation}
\left(\dfrac{\partial y}{\partial\gamma}\right)^{2}=4y^{3}-g_{2}y-g_{3},
\label{eq:Weierstrass}
\end{equation}
with corresponding parameters
\be
g_{2} = \dfrac{a_{2}^{2}}{12}-\dfrac{a_{1}a_{3}}{4}, \quad g_{3}=\dfrac{a_{1}a_{2}a_{3}}{48}-\dfrac{a_{2}^{3}}{216}-\dfrac{a_{0}a_{3}^{2}}{16}.
\ee
Eq.~ (\ref{eq:Weierstrass-sol}) is of elliptic type and its solution is well known, it is given by the Weierstrass function
\begin{equation}
y(\gamma)=\wp(\gamma-\gamma_{in}^{\prime}\,;\, g_{2},\, g_{3})\label{eq:Weierstrass-sol},
\end{equation}
and hence, the $\tilde r$-equation of motion, Eq.~(\ref{eq:Ec cuarto orden}), has the solution
\begin{equation}
\tilde{r}=\dfrac{a_{3}}{4\wp(\gamma-\gamma_{in}^{\prime}\,;\, g_{2},\, g_{3})-\dfrac{a_{2}}{3}}.
\label{eq:sol- weierstrass}
\end{equation}

\subsubsection{Case $\tilde b \to \infty$ }

\begin{center}
\begin{figure}[pbth]
 \includegraphics[width=8cm]{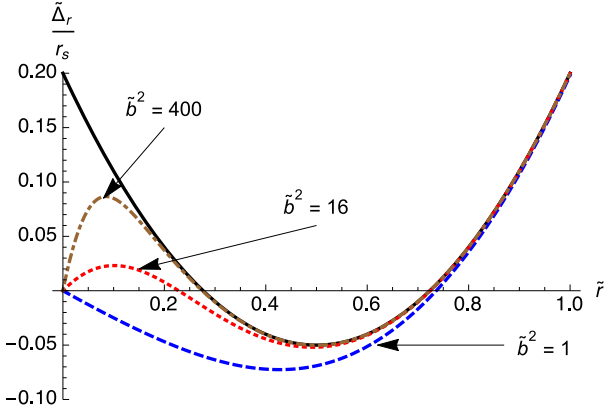}
\caption{The function $\tilde \Delta_r$ for different values of the dimensionless parameter $\tilde b := r_s b$; the solid line corresponds to the RN spacetime with $\tilde Q = \sqrt{0.2}$.} 
\label{fig11}
\end{figure}
\end{center}

In this case we proceed as above but before doing we must analyse the behaviour of the function $\tilde{\Delta}_r$. Fig.~\ref{fig11} shows this function for small and large values of the parameter $\tilde b := r_s b$ and $\tilde Q = \sqrt{0.2}$. 

It can be remarked that for $\tilde b$ large and strictly positive values of $\tilde r$, the function $\tilde \Delta_r$ has a similar behaviour as that of a quadratic polynomial on $\tilde r$ with two real roots. We write then the expression
\begin{equation}
\tilde{\Delta}_{r}=(\tilde{r}-r_{h_-})(\tilde{r}-r_{h_+}),
\label{eq:gtt aprox b grande}
\end{equation}
which corresponds to a quadratic polynomial, its roots being located at the inner and outer event horizons $r_{h_-}$ and $r_{h_+}$ respectively. Using this expression we have
\begin{equation}
\left(\dfrac{\partial\tilde{r}}{\partial\gamma}\right)^{2} = \tilde{r}^{4}E^{2} + 2\tilde{r}^{3}\Delta_{q} + \tilde{r}^{2}\Delta_{q}^{2} - (\tilde{r} - r_{h_-})(\tilde{r} - r_{h_+})(k + \delta\tilde{r}^{2}),
\end{equation}
or equivalently
\be
\left(\dfrac{\partial\tilde{r}}{\partial\gamma}\right)^{2}=a_{0}+a_{1}\tilde{r}+a_{2}\tilde{r}^{2}+a_{3}\tilde{r}^{3}+a_{4}\tilde{r}^{4},
\ee
where
\begin{eqnarray}
&&a_{0} = -k\, r_{h_-} r_{h_+}, \qquad a_{1} = -k(r_{h_-} + r_{h_+}),
\nonumber \\[4pt]
&&a_{2}=\left(\Delta_{q}^{2}-k-\delta\, r_{h_-} r_{h_+}\right), 
\nonumber \\[4pt]
&&a_{3}=[2E\Delta_{q}-\delta(r_{h_-} + r_{h_+})],
\nonumber \\[4pt]
&&a_{4}=(E^{2}-\delta).
\end{eqnarray}
We now proceed as before. First, we make the change of variable $\tilde{r}=\pm\frac{1}{x}+\tilde{r}_{R}$, where $\tilde{r}_{R}$ is one of the roots of the quartic polynomial 
\be
a_{0}+a_{1}\tilde{r}+a_{2}\tilde{r}^{2}+a_{3}\tilde{r}^{3}+a_{4}\tilde{r}^{4}=0,
\ee
and in consequence we arrive to
\be
\left(\dfrac{\partial x}{\partial\gamma}\right)^{2} = b_{0}+b_{1}x+b_{2}x^{2}+b_{3}x^{3},
\ee
where
\begin{eqnarray}
&&b_0 = a_4, \qquad b_1 = a_3 + 4 a_4 \tilde r_R,
\nonumber \\[4pt]
&&b_2 = a_2 + 3 a_3  \tilde r_R + 6 a_4  \tilde r_R^2,
\nonumber \\[4pt]
&&b_3 = a_1 + 2 a_2 \tilde r_R + 3 a_3 \tilde r_R^2 + 4 a_4 \tilde r_R.
\end{eqnarray}
The further change of variable
\be
x=\left(4y-b_{2}/3\right)/b_{3},
\ee
allows us to obtain
\be
\left(\dfrac{\partial y}{\partial\gamma}\right)^{2}=4y^{3}-g_{2}y+g_{3},
\label{weierstrasseq2}
\ee
where
\be
g_{2}=\dfrac{b_{2}^{2}}{12}-\dfrac{b_{1}b_{3}}{4}, \qquad g_{3}=\dfrac{b_{1}b_{2}b_{3}}{48}-\dfrac{b_{2}^{3}}{216}-\dfrac{b_{0}b_{3}^{2}}{16}.
\ee
Eq.~(\ref{weierstrasseq2}) is again Weierstrass' differential equation and therefore we can write immediately its solution
\be
\tilde{r}=\pm\dfrac{b_{3}}{4\wp(\gamma-\gamma_{in}^{\prime}\,;\, g_{2},\, g_{3})-\dfrac{b_{2}}{3}}+\tilde{r}_{R},
\ee
as the solution to the $\tilde r$-equation of motion in this case.

\subsection{Solution of the $\theta(\gamma)-$equation }

The solution of Eq. (\ref{eq:Geodesic-cos-Theta}) with $a<0$ and $D>0$ can be obtained in a straightforward way and it is given by the elementary function
\begin{equation}
\theta(\gamma)=\arccos\left(\dfrac{1}{2a}(\sqrt{D}\sin(\sqrt{-a}\gamma-\gamma_{in}^{\vartheta})-b)\right),
\end{equation}
where $\gamma_{in}^{\vartheta}=\sqrt{-a}\gamma_{in}-\arcsin(\frac{\gamma_{in}+b}{\sqrt{D}})$, $\gamma_{in}$ is the initial value of $\gamma$ and $D := 4 k \kappa$.

\subsection{Solution of the $\phi(\gamma)-$equation}

Eq.~(\ref{eq:Geod=0000E9sica-L}) can be simplified by using first Eq.~(\ref{eq:Theta-Geod=0000E9sica 2-2}) and the change of variable $\xi=\cos\theta$. We have thus
\begin{equation}
d\phi=-\dfrac{d\xi}{\sqrt{\Theta_{\xi}}}\dfrac{\tilde{L}}{1-\xi^{2}}-\dfrac{\xi d\xi}{\sqrt{\Theta_{\xi}}}\dfrac{\Delta_{g}}{1-\xi^{2}},
\end{equation}
where $\Theta_{\xi}$ is given in Eq.~(\ref{eq:Geodesic-cos-Theta}). The resulting equation can be easily integrated and the solution for $a<0$ and $D>0$ is given by
\begin{equation}
\phi(\gamma)=\left.\dfrac{1}{2}(I_{+}+I_{-})\right|_{\xi_{in}}^{\xi(\gamma)}+\phi_{in},
\end{equation}
where
\begin{eqnarray}
I_{\pm} &:=& -sgn(\tilde L \pm \Delta_g) \arcsin \frac f{\sqrt{D}}, 
\nonumber \\[4pt]
&& 
\nonumber \\[4pt]
f &:=& \dfrac{k+\kappa-(\tilde{L}\pm\Delta_{g})^{2}\mp(k+\kappa+(\tilde{L}\pm\Delta_{g})^{2})\xi}{\xi\mp1}.
\end{eqnarray}
Here $sgn(z)$ means the sign function.

For the special case $k=\tilde{L}^{2}$ and $\tilde{L}=\pm\Delta_{g}$, the solution reduces to the simple form 
\begin{equation}
\phi(\gamma)=\left.\dfrac{1}{2} \left( sgn(\tilde L) \arcsin\dfrac{1\pm3\xi}{\xi\mp1} \right) \right|_{\xi_{in}}^{\xi(\gamma)}+\phi_{in},
\end{equation}
where $\phi_{in} := \phi(\gamma_{in})$.  The $\theta$- and $\phi$-motions are actually the same as those obtained by Grunau and Kagramanova~\cite{Grunau:2010gd}.

\section{Conclusions}

In this work we have analysed the geodesic motion of charged test particles for the EBI spacetime. We have seen that even though BI electrodynamics has a more complicated Lagrangian, the orbits followed by massive particles admit a decomposition similar to that of the RN spacetime. 

The main difference between these two spacetimes is encoded into the function $\tilde \Delta_r$, which in the EBI case involves a hypergeometric function. This has a non-trivial influence on the motion of particles.

As seen from Figs.~\ref{fig7}-\ref{fig10}, the RN effective potential is closer in appearance to the typical EBI effective potential, however some differences are noticeable depending on the value of the parameter $k$. In particular from Figs.~\ref{fig7} and~\ref{fig10}, we can deduce that a charged particle can fall to the origin in the EBI case since there is no barrier wall near the origin, meanwhile this will not happen in the RN scenario since there is always a barrier wall for small values of $\tilde r$. Obviously in the Schwarzschild case a massive particle can fall to the origin but in that case there is not electric or magnetic charge involved.

Another note-worthy feature from the geodesic motion in EBI spacetime can be inferred from Figs.~\ref{fig8} and its mirror image Fig.~\ref{fig9}. In the RN spacetime, there are some values of $V_{eff}$ in the interval 4.6-4.8 for which a closed orbit is possible. These orbits disappear in the EBI case, being replaced by unbounded trajectories. Furthermore, for some orbits a turning point in the RN spacetime can be closer to the origin than the corresponding one in EBI spacetime and viceversa. 

We have also analysed two extremes cases where analytical results can be obtained. In both situations the radial geodesic equation of motion is amenable, after a series of transformations, to the differential equation satisfied by the Weierstrass function. In this way a full explicit solution is obtained. 

\acknowledgments

The authors acknowledge support from CONACyT-DFG Collaboration Grant 147492. Daniel Martínez was also supported by CONACyT Fellowship 317495.


\end{document}